\documentclass[twocolumn,english,prl]{revtex4-1}
\usepackage{xcolor}
\usepackage{graphicx}
\usepackage[T1]{fontenc}
\usepackage{float}
\usepackage{textcomp}
\usepackage{amsmath}
\usepackage{amssymb}
\usepackage{graphicx}
\usepackage{esint}
\usepackage{natbib}
\usepackage{color}

\begin{document}

\title{Paired electron motion in interacting chains of quantum dots}


\author{Bart\l{}omiej Szafran}

\affiliation{AGH University of Science and Technology, Faculty of Physics and
Applied Computer Science,\\
 al. Mickiewicza 30, 30-059 Krak\'ow, Poland}

\begin{abstract}
We study the motion of a pair of electrons along two separate parallel chains of quantum dots.
The electrons that are released from the central dot of each chain tend to accompany and not avoid each other.
The correlated electron motion involves entanglement of the wave functions which is generated in time upon 
release of the initial confinement. Observation of the simultaneous presence of  electrons at the same
side of the chain can provide fingerprint of the paired electron motion. 
\end{abstract}

\maketitle

Single-electron charge dynamics in semiconductor double quantum dots is extensively studied in the context of quantum states control \cite{hay,pet,ste,cao,wan} and quantum information processing \cite{dov,cao,pet2}. 
Observation of charge oscillations  from one dot to the other in the time domain is used to evaluate the coherence and energy relaxation times \cite{hay,pet,pet2,cao,wan,ste}.
Shuttling of single electrons across arrays of quantum dots have recently been
performed \cite{fle,mil,bay}. Devices with systems of multiple quantum dots that are mutually coupled by the Coulomb interaction are considered for conditional operations 
on charge \cite{li} and spin \cite{ivan} qubits as well as for wave function entanglement  \cite{shu,sri,sta}.

In this paper we consider two chains of quantum dots each containing a single electron 
and the quantum dynamics  upon release of the initial potential localizing both electrons in the central dots of the chain. We find that the electrons tend to accompany each other in their motion across the chain of dots and that the correlation of the electron motion due to the entanglement of the wave function can be read-out by simultaneous detection of electrons at the ends of the chains.

\begin{figure}[htbp]
\begin{tabular}{l}
 \includegraphics[width=0.5 \columnwidth]{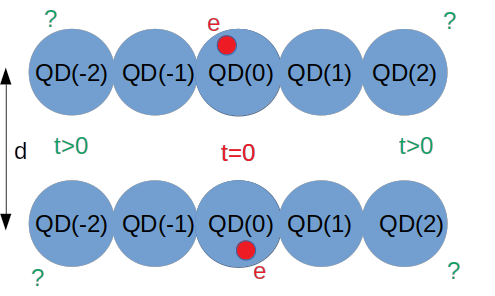} \put(-10,58){ a)} \\
 \includegraphics[width=.7 \columnwidth]{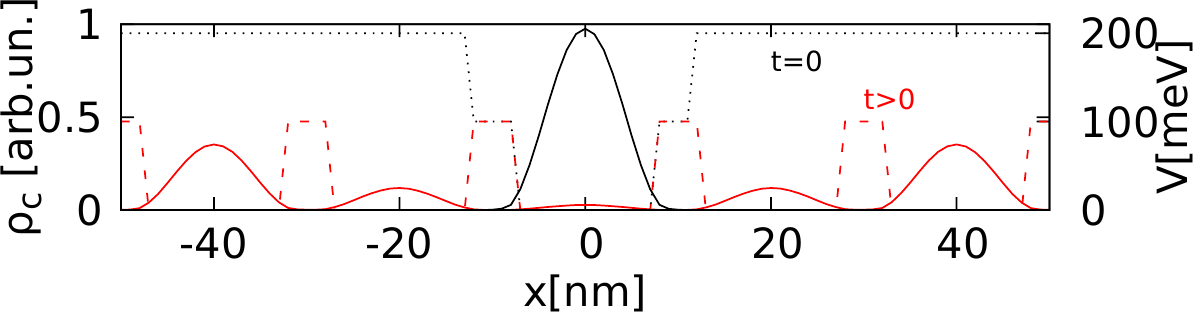} \put(-30,48){b)}
  \end{tabular}
\caption{ (a) Schematics of two chains of five quantum dots 
separated by a distance $d$. A single electron is confined in each chain. 
For the initial condition the electrons are set in the central quantum dots of each chain.
For $t>0$ the confinement is released and the electrons tunnel to other dots.
We consider simultaneous detection of electrons in the same or opposite ends of the chain. 
(b) The black dotted line shows the potential
along the chain that is used for the initial condition.
The red dashed line shows the potential that is set at $t>0$.
With the solid black (red) line we plot the charge density for the ground
state in the initial condition (in the ground state for the potential set at $t>0$)
for $d=100$ nm.
 \label{sze}}
\end{figure}

\begin{figure}
\begin{tabular}{ll}
 \includegraphics[width=0.52 \columnwidth]{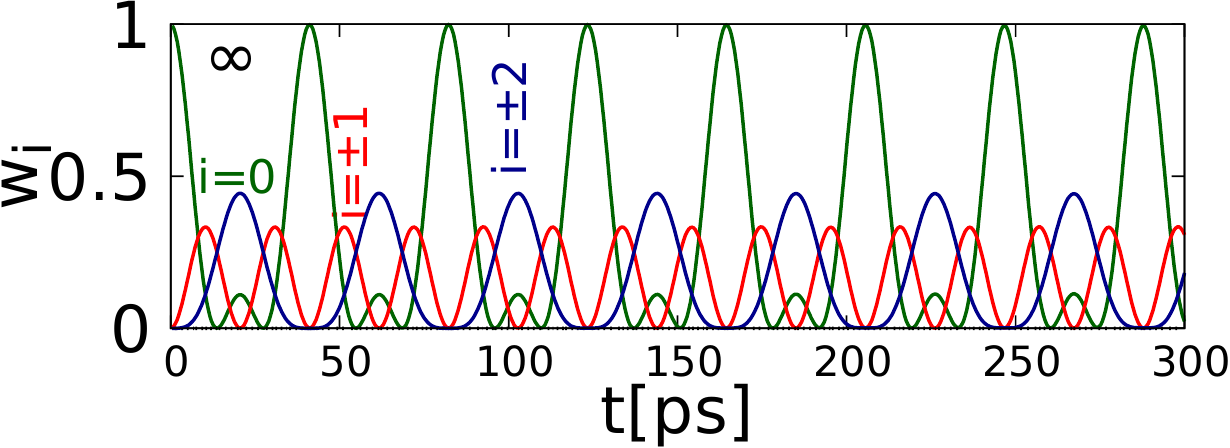} \put(-15,48) {a)} &  \includegraphics[width=0.52 \columnwidth]{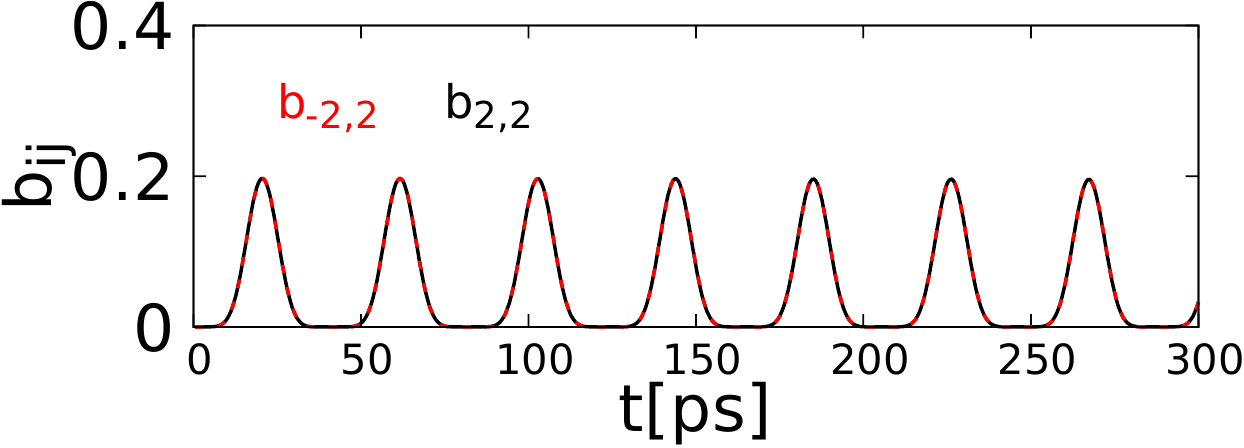}\put(-15,48){ b)}  \\
 \includegraphics[width=0.52 \columnwidth]{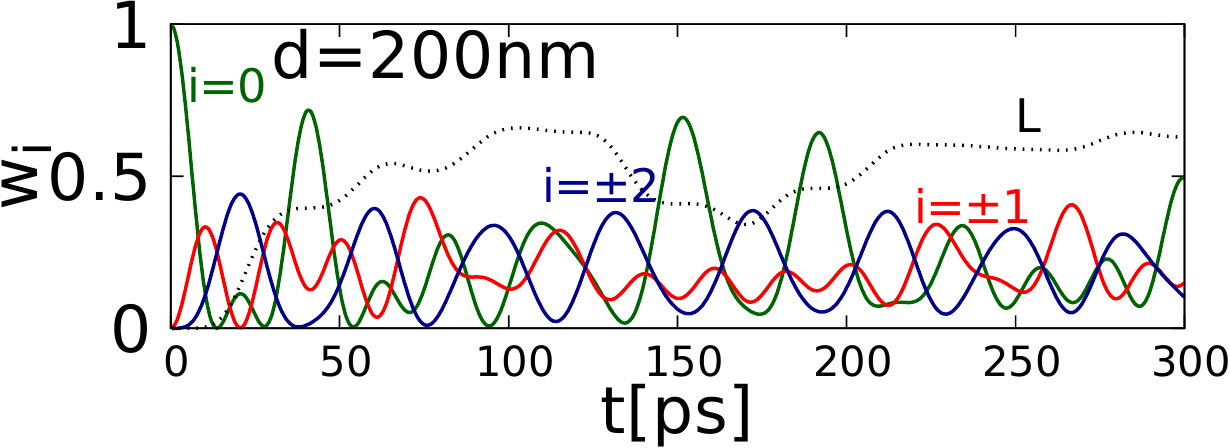}\put(-15,48){ c)}  &  \includegraphics[width=0.52 \columnwidth]{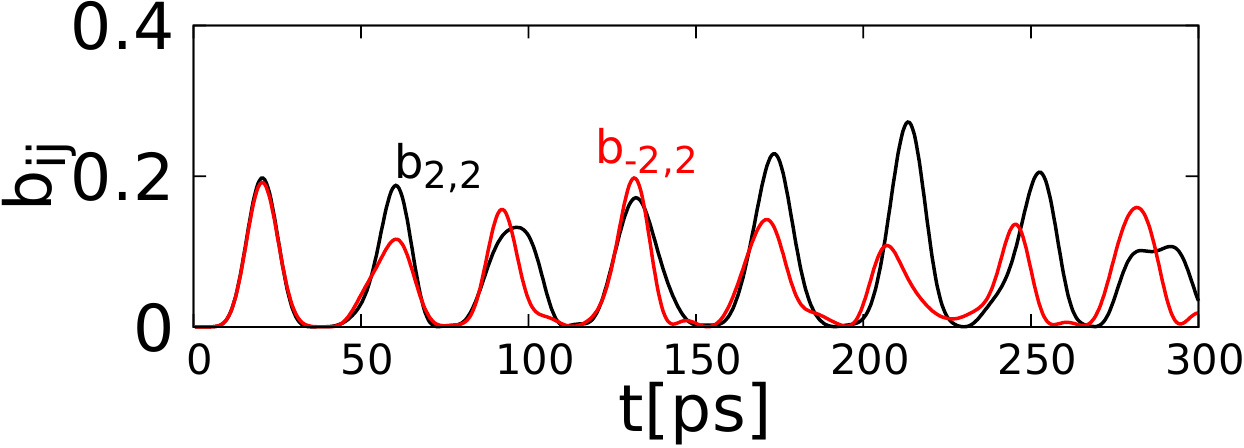} \put(-15,48){ d)} \\
 \includegraphics[width=0.52 \columnwidth]{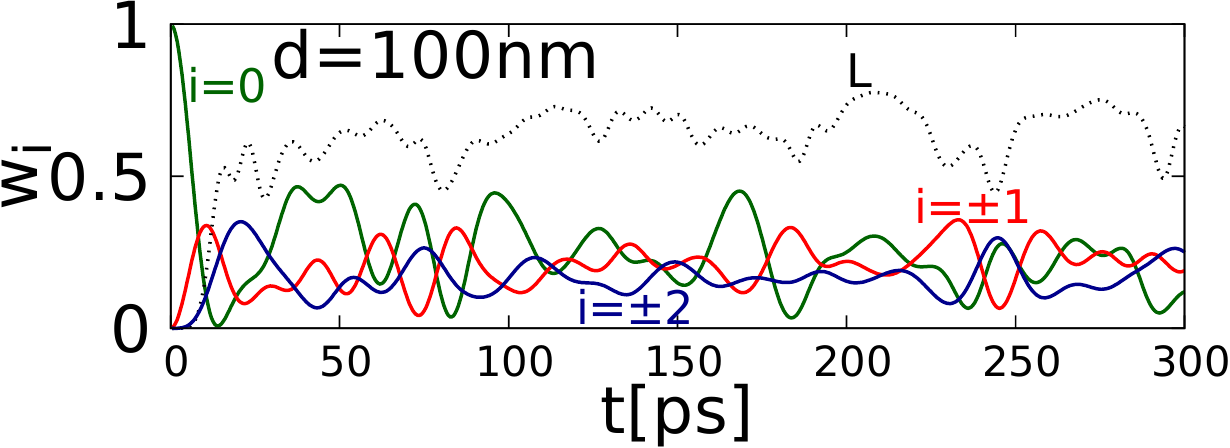} \put(-15,48){ e)} & \includegraphics[width=0.52 \columnwidth]{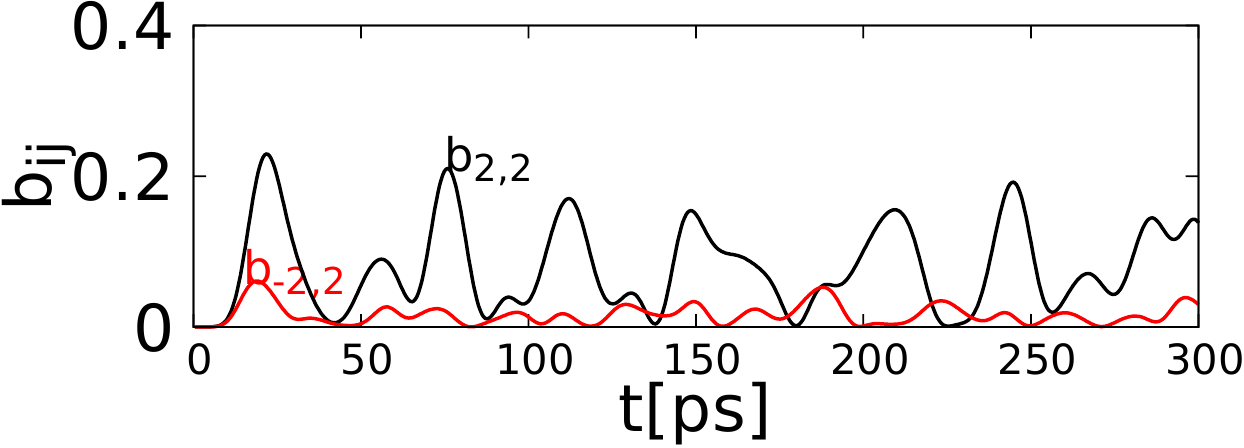}\put(-15,48){ f)}  \\
 \includegraphics[width=0.52 \columnwidth]{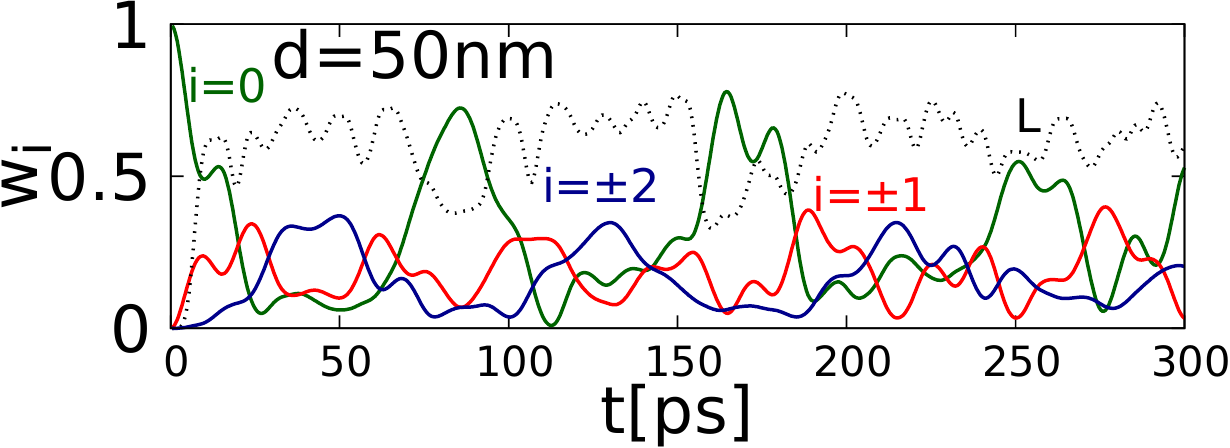} \put(-15,48){ g)} &  \includegraphics[width=0.52 \columnwidth]{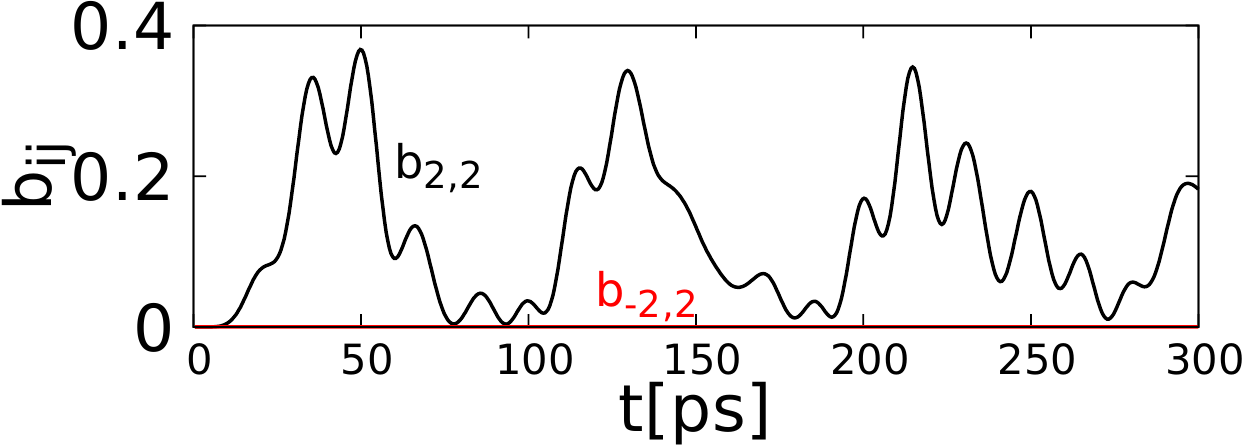} \put(-15,48){ h)} 

\end{tabular}
\caption{Left column: The probability $w_i$ to find an electron in dot $i$ of the chain.  
Right column: probabilities to find electrons in the same $b_{2,2}=b_{-2,-2}$ or opposite
ends of the chain $b_{-2,2}=b_{2,-2}$, for $d=\infty$ (a,b), $d=200$ nm (c,d), $d=100$ nm (e,f) and $d=50$ nm (g,h). In the left column the linear entropy $L$ is plotted with the dotted line. 
 \label{szex}}
\end{figure}

\begin{figure}[htbp]
\begin{tabular}{lll}
 \includegraphics[width=0.3 \columnwidth]{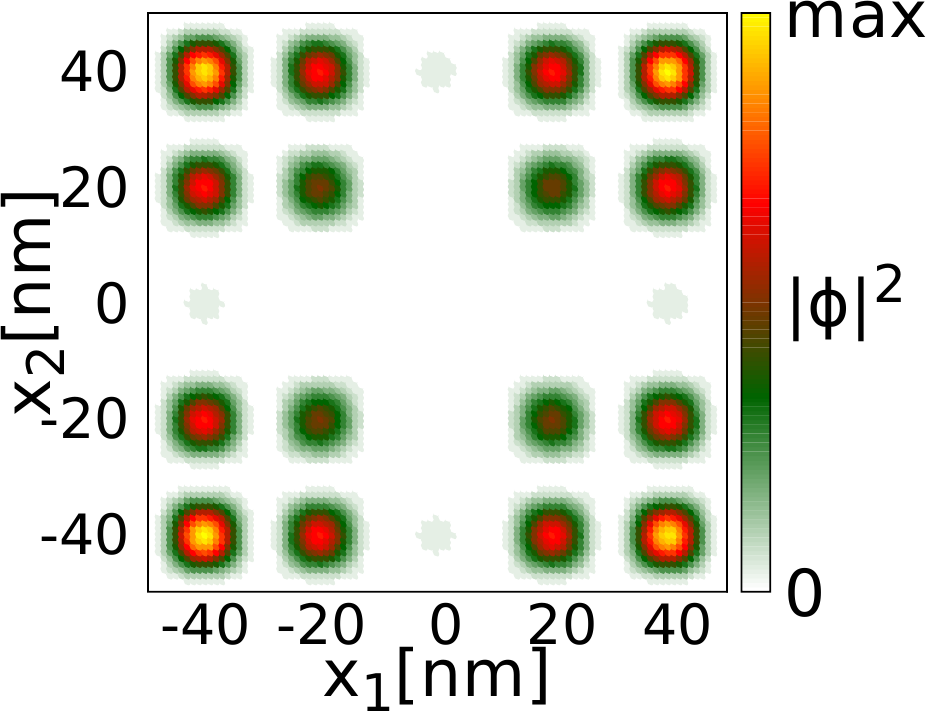} \put(-32,62){ a)} &  \includegraphics[width=0.3 \columnwidth]{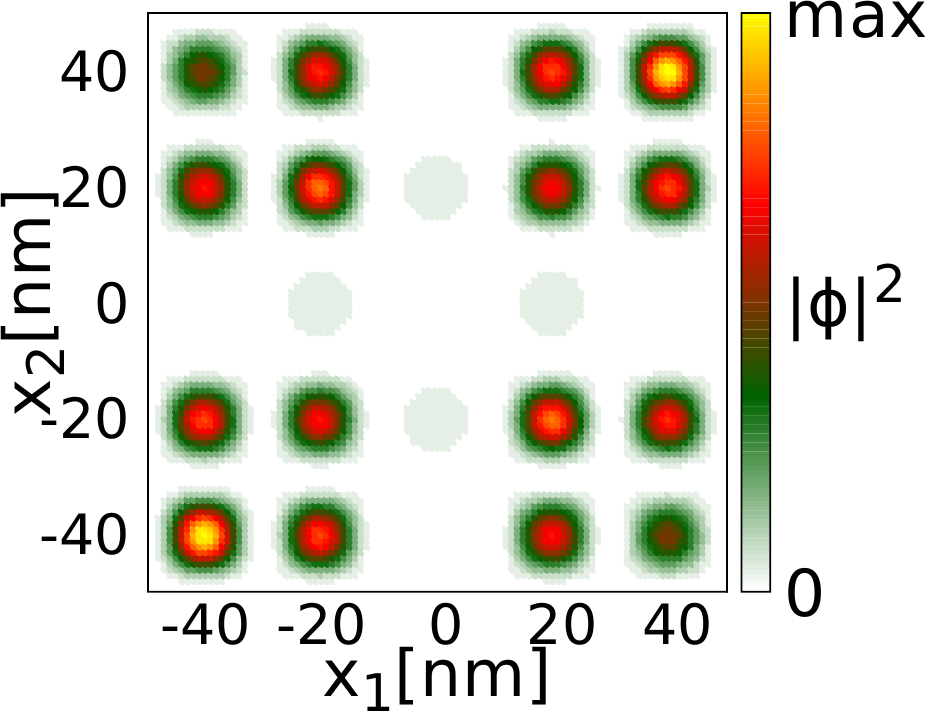}\put(-32,62){ b)}  & \includegraphics[width=0.3 \columnwidth]{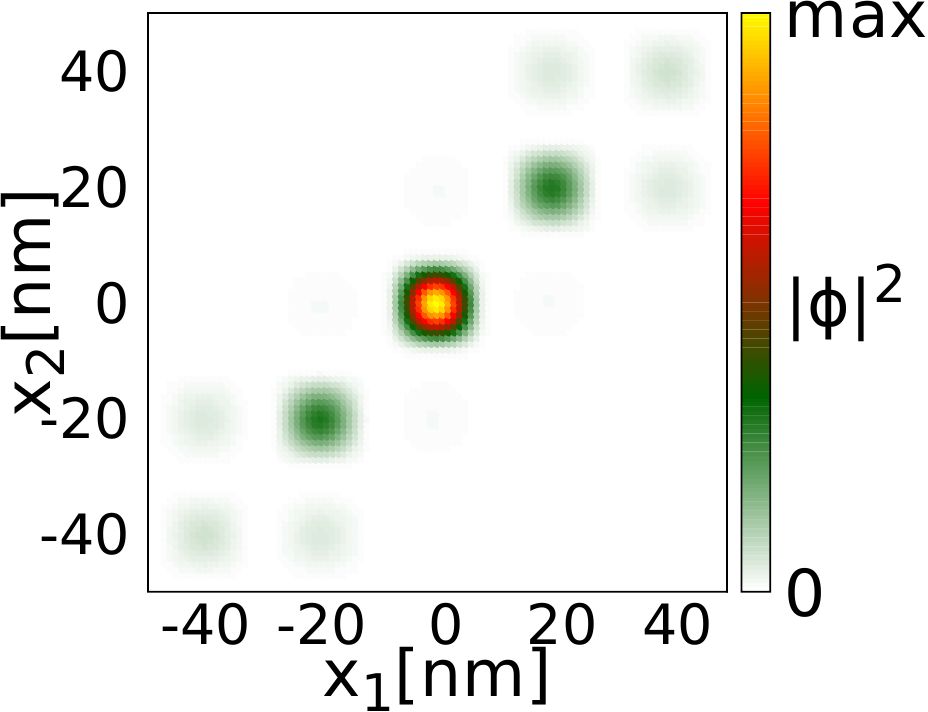}\put(-32,62){ c)}\\

 \includegraphics[width=0.3 \columnwidth]{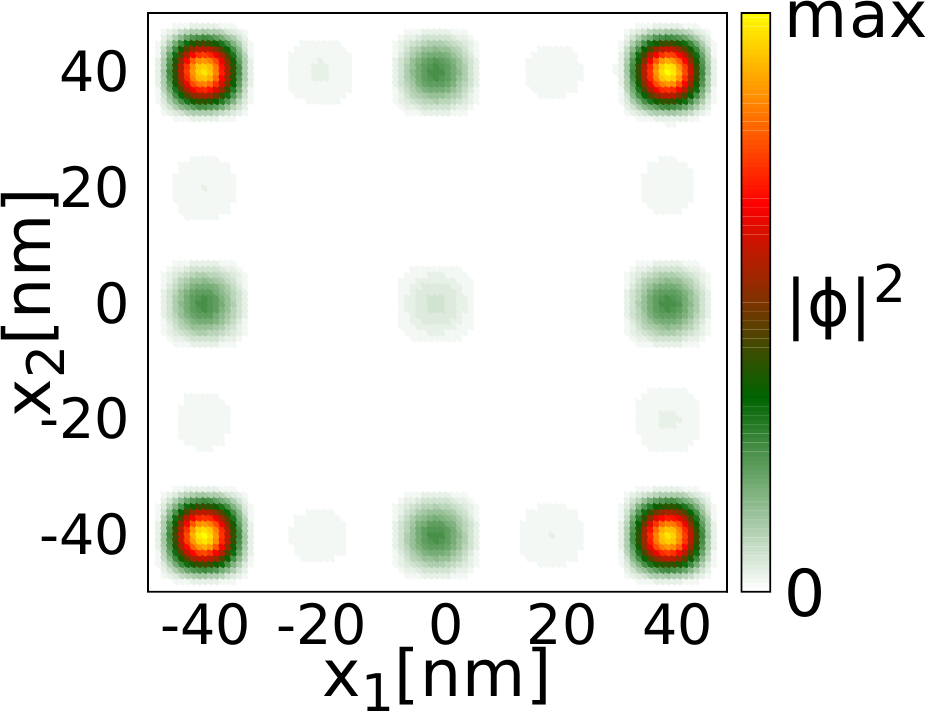} \put(-32,62){d)} &  \includegraphics[width=0.3 \columnwidth]{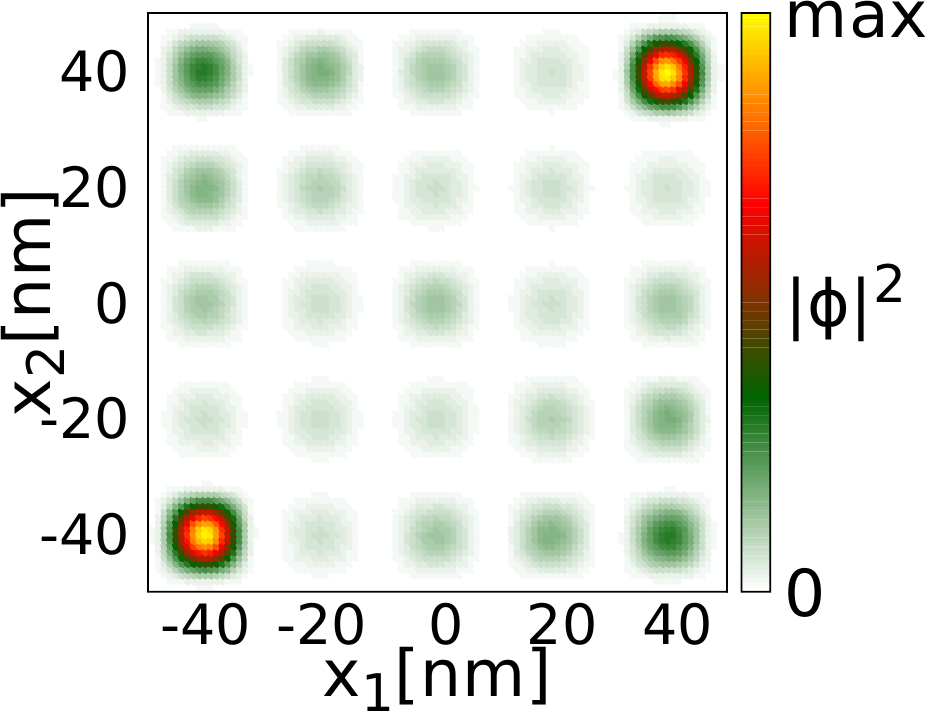}\put(-32,62){ e)}  & \includegraphics[width=0.3 \columnwidth]{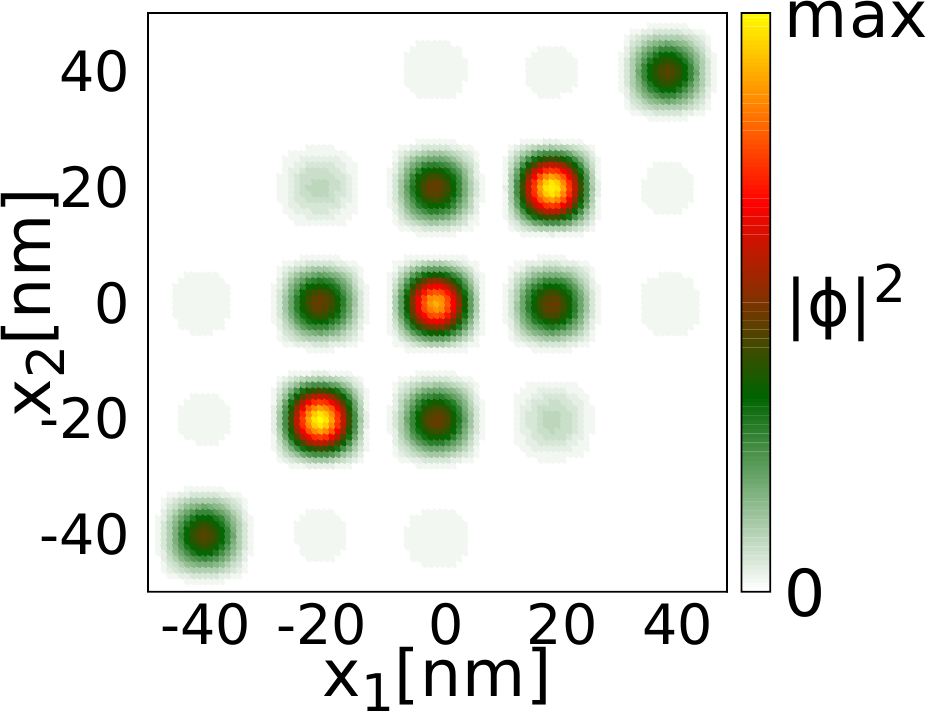}\put(-32,62){ f)}\\

 \includegraphics[width=0.3 \columnwidth]{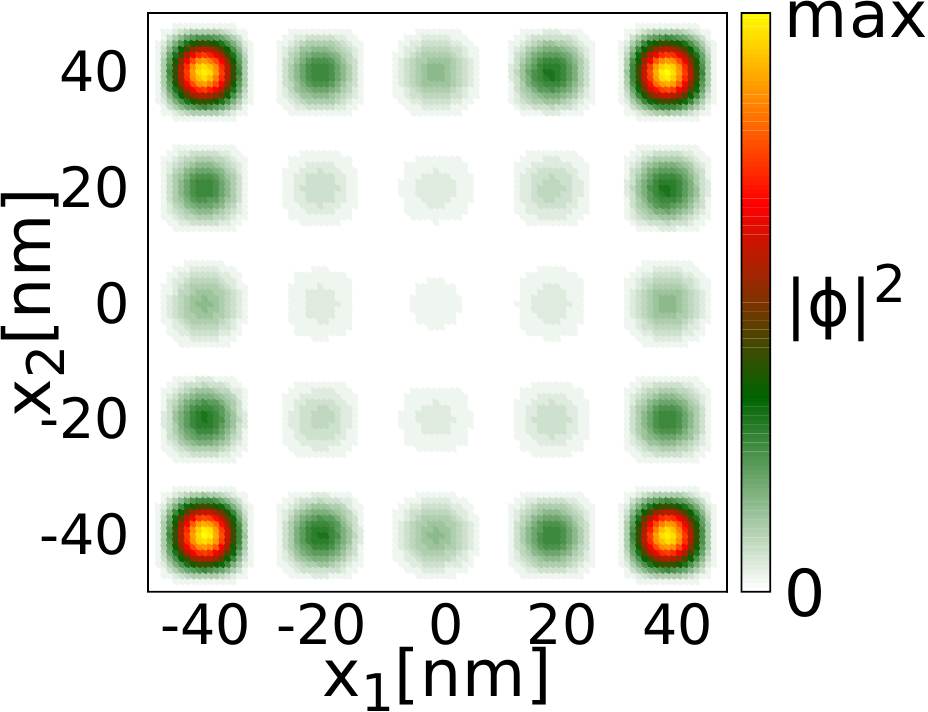} \put(-32,62){ g)} &  \includegraphics[width=0.3 \columnwidth]{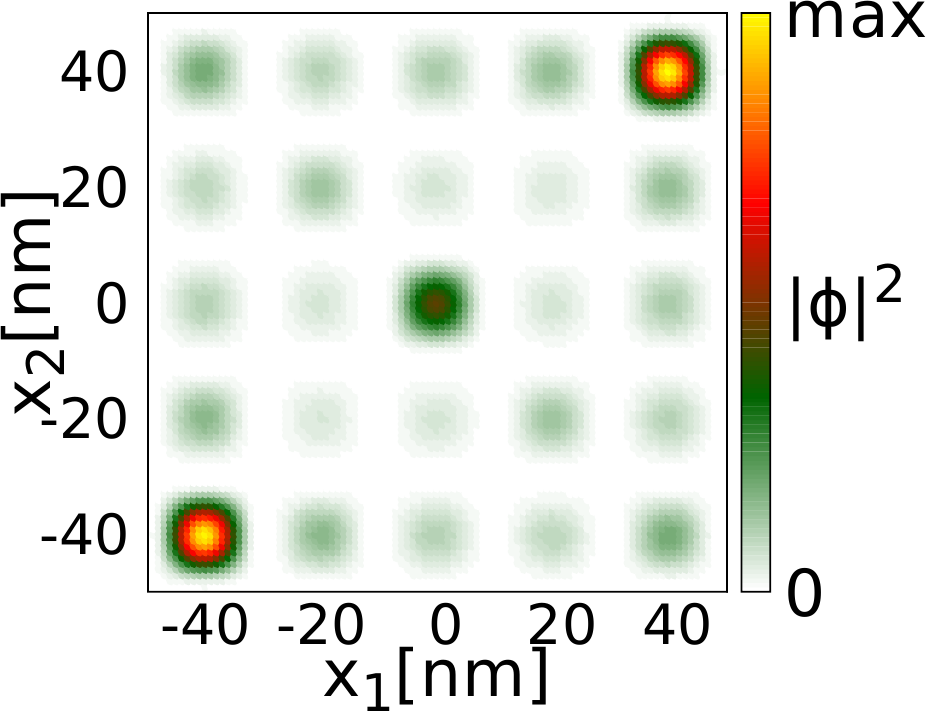}\put(-32,62){ h)}  & \includegraphics[width=0.3 \columnwidth]{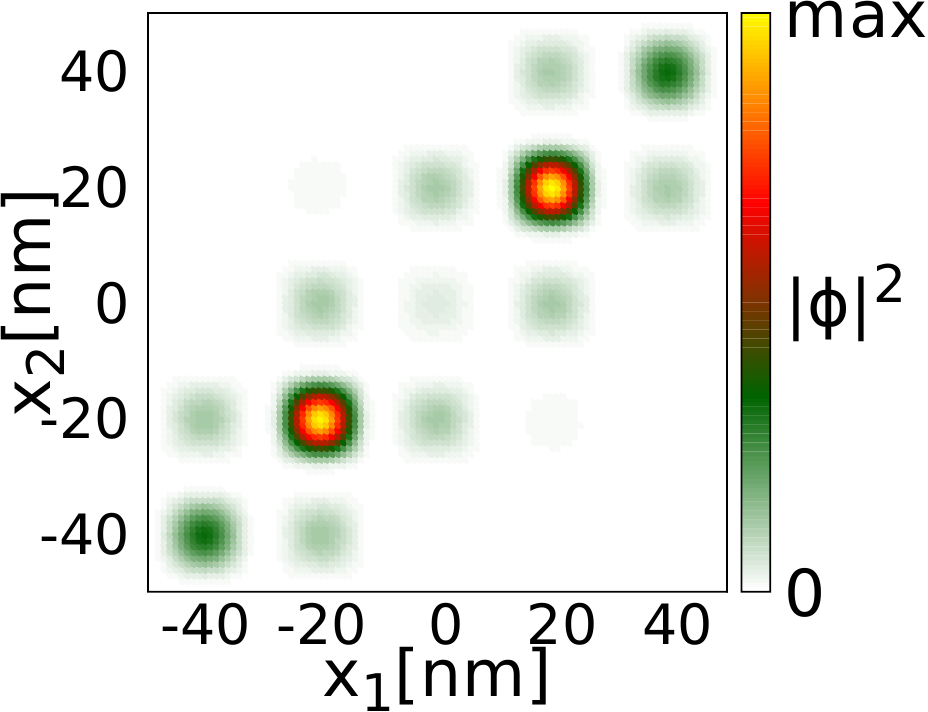}\put(-32,62){ i)}

\end{tabular}
\caption{ Probability densities on the $x_1,x_2$ plane at 
$t=14.7$ ps (a-c), $t=19.6$ ps  (d-f) and  $t=24.5$ ps  (g-i)  upon release 
of the initial potential for $d=\infty$ (a,d,g), $d=100$ nm (b,e,h), and $d=50$ nm (c,f,i).
 \label{sze}}
\end{figure}

\begin{figure*}[htbp]
\begin{tabular}{ll}
 \includegraphics[width=0.6 \columnwidth]{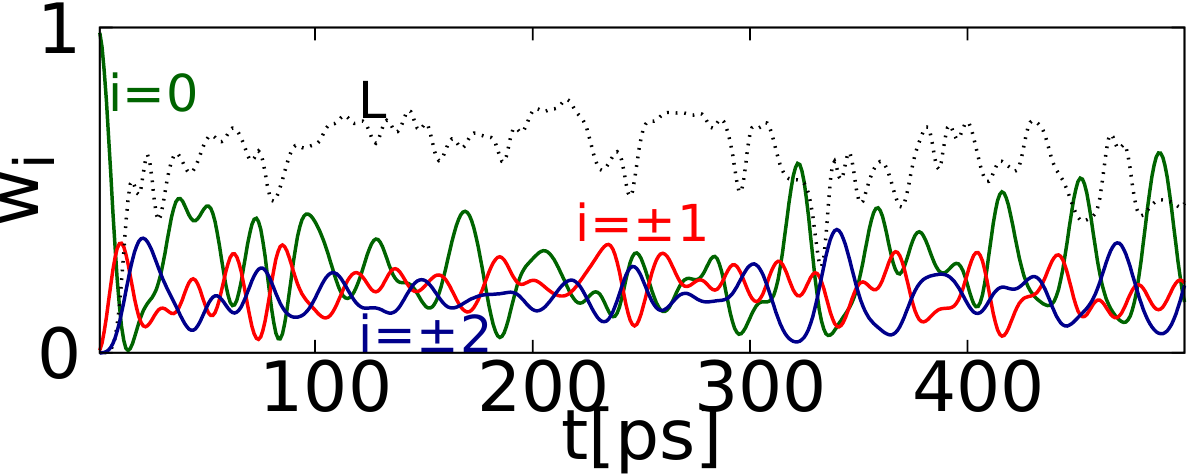}\put(-108,64){ a)} \put(-38,64) {\rotatebox{0}{\small $\alpha=0.005, T_1=1.35$ns}} &  \includegraphics[width=0.7 \columnwidth]{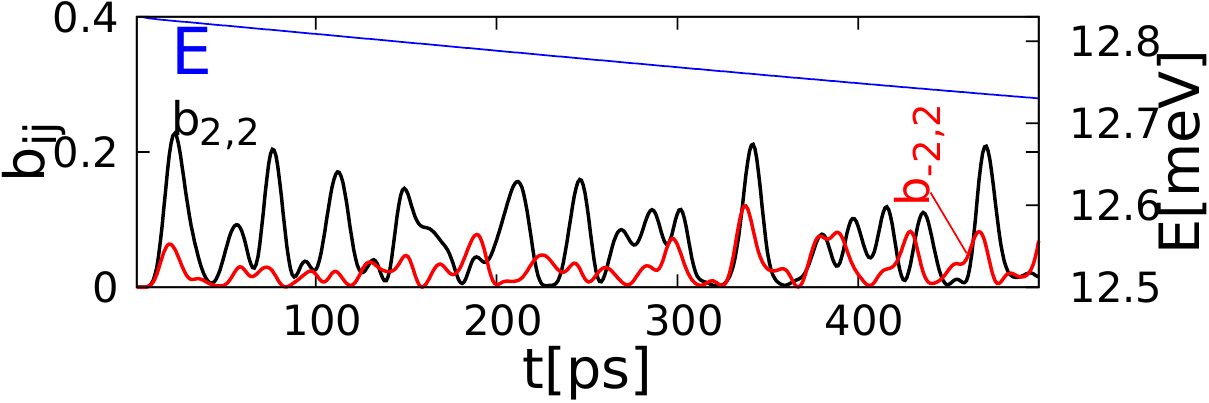} \put(-58,64){ b)} \\
 \includegraphics[width=0.6 \columnwidth]{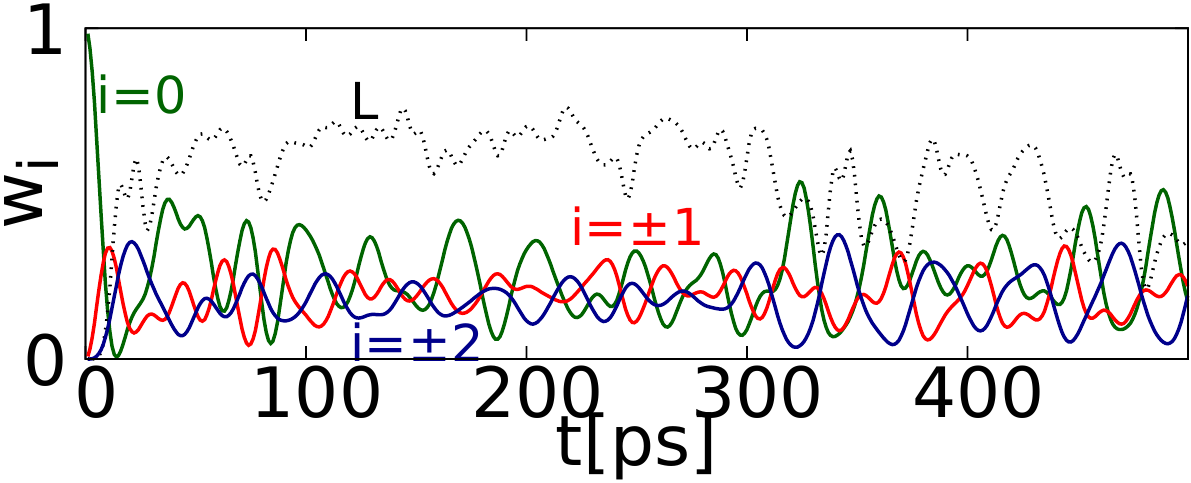} \put(-108,64) {c)}\put(-38,64) {\rotatebox{0}{\small $\alpha=0.01, T_1=680$ps}} &  \includegraphics[width=0.7 \columnwidth]{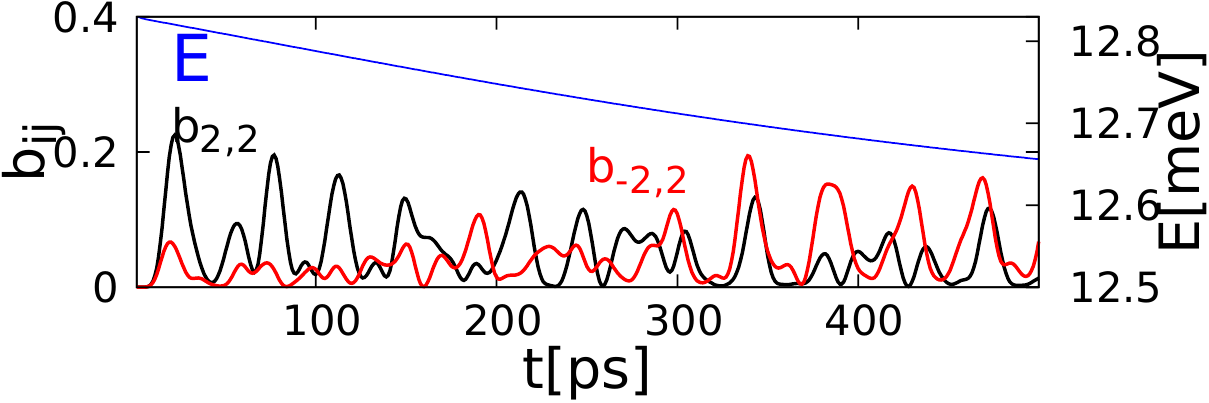}\put(-58,64){ d)}  \\
 \includegraphics[width=0.6 \columnwidth]{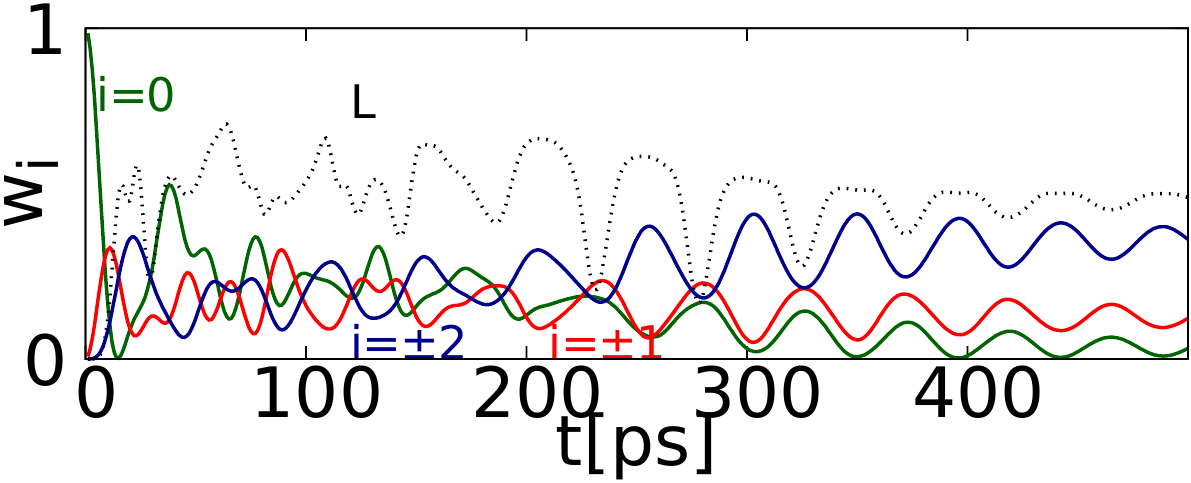}\put(-108,64){ e)} \put(-38,64) {\rotatebox{0}{\small $\alpha=0.05, T_1=136$ps}} &  \includegraphics[width=0.7 \columnwidth]{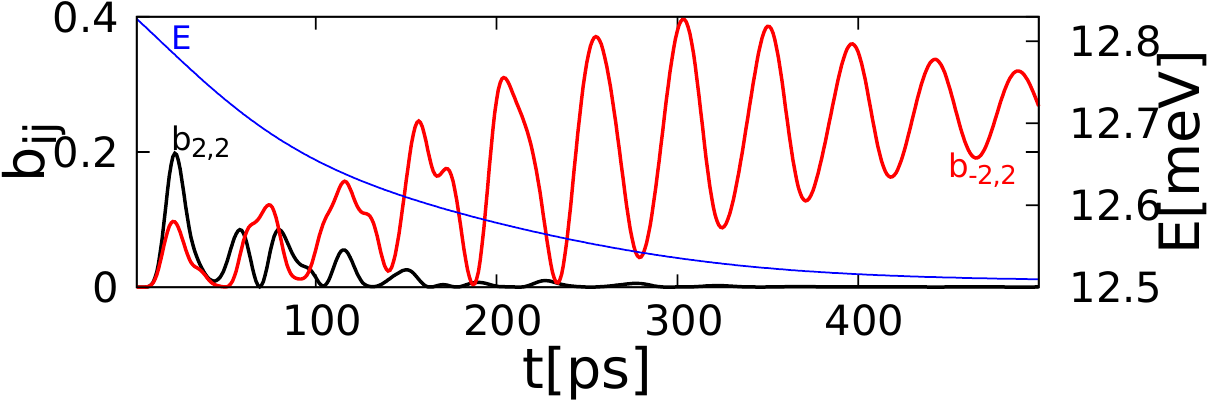} \put(-58,64){ f)} \\
 \includegraphics[width=0.6 \columnwidth]{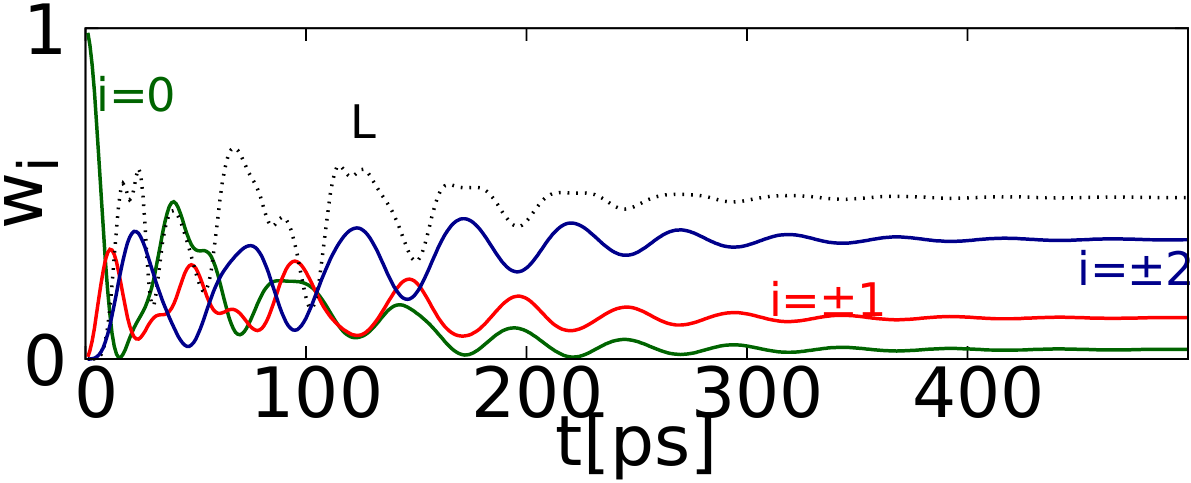} \put(-108,64){ g)}\put(-38,64) {\rotatebox{0}{\small $\alpha=0.1, T_1=68$ps}} & \includegraphics[width=0.7 \columnwidth]{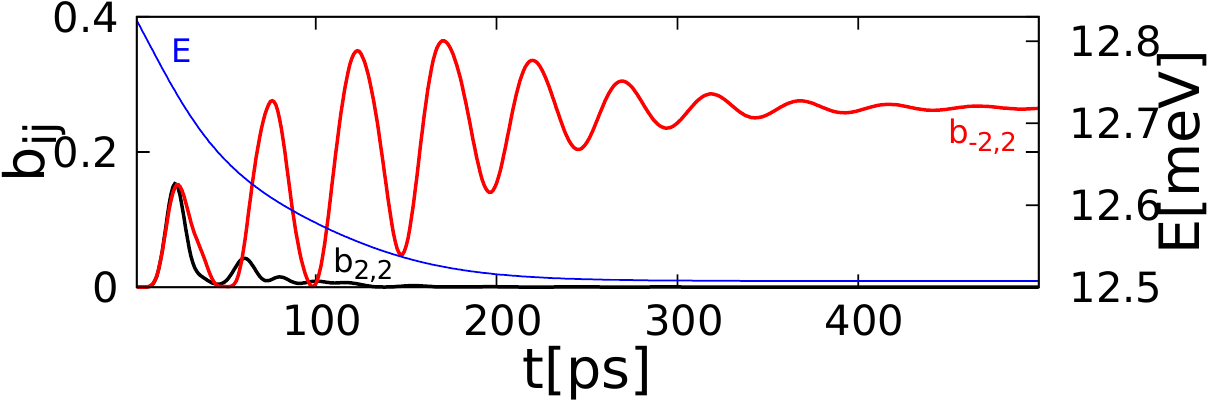}\put(-58,64){ h)}  
\end{tabular}
\caption{
Left column: The probability $w_i$ to find an electron in dot $i$ of the chain
for the inter-chain distance of  $d=100$ nm with the energy relaxation introduced by the
imaginary time stepping. 
Right column: probabilities to find electrons in the same $b_{2,2}=b_{-2,-2}$ or opposite
ends of the chain $b_{-2,2}=b_{2,-2}$. In the left column the linear entropy $L$ is plotted with the dotted line. 
The ratio of imaginary steps to the total number of steps  $\alpha$ is $0.005$ (a,b), $0.01$ (c,d), $0.05$ (e,f) and $0.1$ (g,h), respectively. 
The corresponding result without the relaxation is given in Fig. 2(e,f).
The blue line in (b,d,f,h)
shows the expectation value of the energy. The effective relaxation time $T_1=1.36$ ns (a-b), 680 ps (c-d), 136 ps (e-f) 
and 68 ps (g-h) is
obtained as a fit to the quantum average of the energy   $E(t)=E_f+(E_i-E_f)\exp(-t/T_1)$, with $E_i$ and $E_f$ standing for the initial and final energy, that correspond to the ground-state energies for the potential set at $t=0$ and at $t>0$, respectively [Fig. 1(b)].  
 \label{rel}}
\end{figure*}

\begin{figure}[h]
 \includegraphics[width=0.5 \columnwidth]{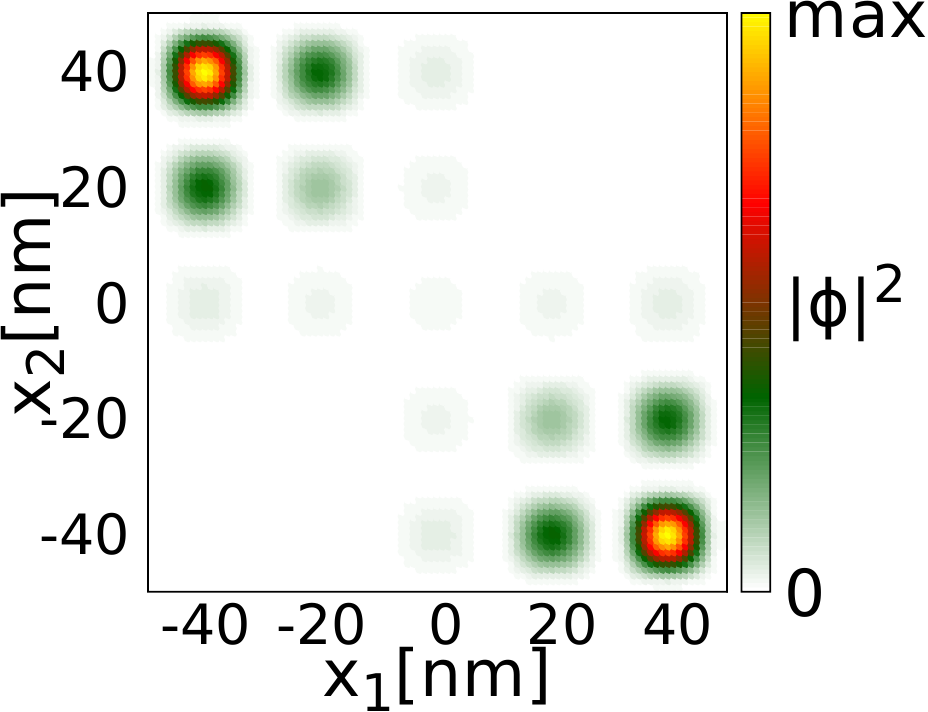}
\caption{Ground-state probability density for the potential introduced at $t>0$ [see Fig. 1(b)]. }
\end{figure}

We work in a model of quasi one-dimensional confinement in two parallel chains of quantum dots [Fig. 1(a)]. The chains are separated by a distance $d$. Each of the chains contains a single electron. We neglect the tunneling between the chains, so that the spin has no influence on the dynamics of the spatial wave functions. We assume that the interaction does not affect the state of quantization in the transverse direction and apply a model of one-dimensional confinement in each chain. The Hamiltonian is taken in form \begin{equation} H=-\sum_{i=1}^2 \frac{\hbar^2}{2m}\frac{\partial^2}{\partial x_i^2}+V(x_i,t)+\frac{e^2}{4\pi\epsilon\epsilon_0 |r_{12}|},\end{equation}
 with $r_{12}=\sqrt{d^2+(x_1-x_2)^2}$, and material parameters corresponding to Si, i.e. the electron effective band mass $m=0.2m_0$, with $m_0$ standing for the electron mass in vacuum, and the dielectric constant $\epsilon=12$.
The potential $V(x,t)$ is taken in the same form for both the chains. We use the finite difference approach for the wave function $\phi(x_1,x_2,t)$ spanned on a grid with mesh spacing of $\Delta x=0.5$ nm. 

As the initial condition we take the ground-state of the electron pair
for the potential that confines the electrons to the central dots of the chain
[see the dotted black line in Fig. 1(b)]. For $t>0$ the potential is changed: $V$ is lowered for other dots 
to the level of the central one [red dashed line in Fig. 1(b)]. Upon  release of the initial confinement the electrons tunnel  across the barriers that are 100 meV high and 6 nm wide
and separate quantum dots of length of 14 nm.

In order to describe the dynamics of the system we solve the two-electron Schr\"odinger equation $i\hbar\frac{\partial \phi}{\partial t}=H\phi$ to account for the electron motion using the Askar-Cakmack \cite{asca} explicit scheme \begin{equation} \phi(x_1,x_2,t+dt)=\phi(x_1,x_2,t-dt)+\frac{2dt}{i\hbar} H\phi(x_1,x_2,t)\end{equation}  with the time step $dt$  equal to half the atomic time unit. The solution is exact in the numerical sense with a full account taken for the electron-electron correlation. 

In the left column of Fig. 2 we plot the probability $w_i$ to find an electron in the dot $i$ as a function of time
\begin{equation}
w_i(t)=\int_{qd_i} dx_1\int_{-\infty}^\infty  dx_2 |\phi(x_1,x_2,t)|^2,
\end{equation}
where $qd_i$ stands for integration over the i-th quantum dot.
The $w_i$ probabilities are the same for both the chains.
The probabilities of simultaneous presence of electrons in the quantum dot $i$ and $j$ of the respective chains are calculated as
\begin{equation}
b_{ij}(t)=\int_{qd_i} dx_1\int_{qd_j} dx_2 |\phi(x_1,x_2,t)|^2,
\end{equation}
In the right column of Fig. 2 we plot the
 probability of simultaneous electron detection  at the extreme dots located at the ends of the chains: 
 in the dots at the same ($b_{2,2}=b_{-2,-2}$) or opposite ($b_{-2,2}=b_{2,-2}$) ends of the chains. 

In the absence of the electron-electron interaction ($d=\infty$) the time evolution is periodic [Fig. 2(a)] and the electrons are found with equal probability in the same or opposite ends of the chain [Fig. 2(b)]. 
For finite $d$ the interaction enters the dynamics [Fig. 2(c-h)] and electron motion becomes correlated, with a larger probability to find both electrons at the {\it same} end of the chain.
Counterintuitively, the electrons accompany and do not avoid each other
with $b_{2,2}\gg b_{-2,2}$ for $d=100$ nm [Fig. 2(f)] and $d=50$nm [Fig. 2(h)]. Figure 3 shows the snapshots of the probability density for $d=\infty$  (left column), $d=100$ nm (central column), and $d=50$ nm (right column). The stronger the electron-electron interaction the stronger is the correlation of the electron motion -- the probability density tends to gather at the diagonal of the $x_1,x_2$ plane for small $d$.

The correlation appears due to entanglement of the two-electron wave function. The left column in Fig. 3
calculated for $d=\infty$ corresponds to a separable wave function with the relative probability to find and an electron in dot $x_1$ independent of the position of the electron in $x_2$ chain, with
same values of the probability density on the diagonal $x_1=x_2$ and anti-diagonal ($x_1=-x_2$) of the plot. In presence of the interaction [central and right columns of Fig. 3 for $d=100$ nm and $d=50$ nm] a distinct imbalance of the density at the diagonal and anti-diagonal appears. 

 In order to quantify the entanglement we use the linear entropy \cite{buscemi,amico,linen} $L=1-{\mathrm Tr} \rho^2_r$,
where $\rho^2_r(x_1,x_2)\equiv \int \rho_r(x_1,x_3,t)\rho_r(x_3,x_2,t) dx_3$,   is the square of the reduced density matrix $\rho_r(x_1,x_2,t)\equiv \int \phi^*(x_1,x_3,t)\phi(x_2,x_3,t) dx_3$.
The entropy of separable system is $L=0$ and the maximal value of the entropy for the electron pair is 1 \cite{amico}. 
$L$ is plotted with the dotted lines in the left panel of Fig. 2. The entropy remains zero $L=0$
in the absence of the interaction and it is nearly zero $L\simeq 0$  for the initial state of the interacting pair with the electrons localized in the central dots of the chain. $L$ grows from zero when the electrons are released from the central dot with the confinement change at $t>0$.

The reason for the paired electron motion  in the coherent electron dynamics is the energy conservation. 
In the chain of five quantum dots the five lowest energy levels form a band of width of about 0.2 meV while the higher energy band is about 17 meV higher. The electron-electron interaction for $d=100$ nm in the initial state is  about 1.5 meV. 
The interaction energy is therefor too large to be transferred to the excitations within the lowest energy band and it is too low to excite the higher energy band. In consequence the electrons need to move in pair to conserve the energy and keep  the electron-electron interaction strong.

In order to observe the correlated electron behavior one needs to perform 
simultaneous detection of the electrons at the ends of each chain. The correlated state corresponds to higher electron-electron interaction energy, so that the energy relaxation process with its characteristic time ($T_1$) should be of the principal concern. The correlation instead of anticorrelation can be found only provided that the energy relaxation time ($T_1$)
is longer than the time needed for the electrons to reach the extreme dots of the chain. 
The  relaxation and coherence times in double quantum dots can reach several nanoseconds \cite{pet2,cao} at most.

We simulated the energy relaxation by introducing the imaginary time $dt\rightarrow -idt$ 
to the solution of the time-dependent Schr\"odinger equation. This substitution, known as Wick rotation \cite{wick}  is used in the diffusion Monte Carlo methods \cite{mcm}. The evolution in the imaginary time \cite{itm,itm2} brings the system to the ground-state. In our calculation the rate of the energy relaxation is controlled by the $\alpha$ parameter which indicates the ratio of imaginary time steps 
to the total number of steps.

In Fig. 4 we plotted the results of the electron dynamics including the simulation of the energy relaxation processes
 for the inter-chain distance of $d=100$ nm for varied values of the $\alpha$ parameter.
  In the left column of Fig. 4 we plotted the electron localization within the wells and in the right column the probabilities
to find the electrons at both or opposite ends of the chain. With the blue line in the right column of Fig. 4 we plotted the energy (expectation value of the Hamiltonian). The energy dependence on time is found nearly exponential 
with the relaxation time $T_1$, $E(t)=E_f+(E_i-E_f)\exp(-t/T_1)$, with $E_i$ standing for the energy of the initial state and $E_f$ as the ground-state energy for the electron pair in the potential that is introduced for $t>0$ [Fig. 1(b)]. The corresponding probability density for the ground state is given in Fig. 5.  Probability to find both electrons at the opposite side of the chain exceeds the one for both electrons at the same side of the chain around
half the relaxation time.

An experimental study of the correlation and anticorrelation effects discussed here requires
fabrication of quantum dot arrays with mutual charge coupling.
The quantum dots need to be gated for manipulation of the confinement potential in time.
The arrays or multiple quantum dots connected in series were recently considered  \cite{fle,mil,bay,zajac}, 
and the potential switching is routinely used for discussion of e.g. spin and charge \cite{hay,pet,ste,cao} qubits in quantum dots.  The charge coupling between the gated arrays has been used in experiments
on controlled quantum logic \cite{li,ivan} and entanglement generation \cite{shu}. 
Finally, an experiment calls for the charge detection at the extreme dots of the chain. 
In recent experiments on quantum dot arrays the presence of electrons in chosen quantum dots
of the chain was detected by charge sensors e.g. by a nearby quantum dot \cite{mil,zajac} 
or a quantum point contact \cite{bay}. 

The present model assumed a one-dimensional confinement along the chain of quantum dots ($x$ axis). 
The assumption is justified when interaction energy of electrons localized in separate chains of quantum dots are smaller than the  spatial quantization in the transverse directions ($y,z$). 
The electrostatic quantum dots can be taylored from the two-dimensional electron gas \cite{hay,pet,ste,cao} (2DEG)
for which the state of quantization along the the growth direction ($z$) is frozen by the strong vertical confinement. The confinement energy along the other transverse direction ($y$) needs to surpass the 
interchain interaction energy. Alternatively, chains of quantum dots defined in two separate quantum 
wires can be used instead of the 2DEG systems. The electrostatic quantum dots on gated Si \cite{silicon}, InAs \cite{inas} or InSb \cite{insb}
quantum wires,
as well as on semiconducting carbon nanotubes \cite{rmp} can be considered for this purpose.

In summary, we have studied  the coherent dynamics of system of two interacting electrons, each confined in separate chain of quantum dots, 
 upon release of the confinement potential that initially keeps the electrons in the central dot of the chain.
We find  entanglement generation and a paired  motion of electrons which are correlated instead of anticorrelated and tend to move together 
along the chain.
  The detection of electrons  at the ends of the chains provides fingerprint of the correlation and spatial entanglement of the wave functions. 
The relaxation processes turn correlation into anti-correlation but preserve the entanglement of the wave function.

\section*{Acknowledgments}
This work was supported by the National Science Centre (NCN) according to decision DEC-2016/23/B/ST3/00821
and by the Faculty of Physics and Applied Computer Science AGH UST statutory research tasks within the subsidy of
Ministry of Science and Higher Education. The calculations were performed
on PL-Grid Infrastructure.

\end{document}